\documentclass[pra,letterpaper,aps,10pt,superscriptaddress,twocolumn,floatfix,showpacs, nobibnotes]{revtex4-1}
\usepackage{graphicx}
\usepackage{amsmath}
\usepackage{amsfonts}
\usepackage{comment}
\usepackage{float}
\usepackage{amssymb}
\usepackage{epsfig}
\usepackage{epstopdf}
\DeclareGraphicsExtensions{.pdf,.eps,.png,.jpg,.mps,.svg}
\usepackage[pdftex]{color}
\usepackage{amsmath,graphicx,amssymb,braket,xcolor,subfigure,upgreek}
\usepackage[colorlinks, linkcolor=blue, citecolor=blue, urlcolor=blue, breaklinks=true]{hyperref}
\usepackage{microtype}
\microtypesetup{expansion=false}
\usepackage{bbm}
\usepackage{color}
\usepackage{dsfont}
\usepackage{url}

\usepackage{gensymb}
\usepackage[symbol*]{footmisc}

\DefineFNsymbolsTM{otherfnsymbols}{%
  \textasteriskcentered *
  \textsection   \mathsection
  \textdagger    \dagger
  \textdaggerdbl \ddagger
  \textbardbl    \|%
  \textparagraph \mathparagraph
}%
\setfnsymbol{otherfnsymbols}

\begin{document}

\title{Hybrid THz architectures for molecular polaritonics}
\author{A. Jaber}
\affiliation{Department of Physics, University of Ottawa, Ottawa, ON K1N 6N5, Canada}
\affiliation{Max Planck Centre for Extreme and Quantum Photonics, Ottawa, ON K1N 6N5, Canada}

\author{M. Reitz}
\affiliation{Max Planck Centre for Extreme and Quantum Photonics, Ottawa, ON K1N 6N5, Canada}
\affiliation{Max Planck Institute for the Science of Light,
D-91058 Erlangen, Germany}
\affiliation{Department of Physics, Friedrich-Alexander-Universit\"{a}t Erlangen-N\"urnberg, D-91058 Erlangen, Germany}

\author{A. Singh}
\affiliation{Department of Physics, University of Ottawa, Ottawa, ON K1N 6N5, Canada}
\affiliation{Max Planck Centre for Extreme and Quantum Photonics, Ottawa, ON K1N 6N5, Canada}

\author{A. Maleki}
\affiliation{Department of Physics, University of Ottawa, Ottawa, ON K1N 6N5, Canada}
\affiliation{Max Planck Centre for Extreme and Quantum Photonics, Ottawa, ON K1N 6N5, Canada}

\author{Y. Xin}
\affiliation{Iridian Spectral Technologies Ltd., Ottawa, ON K1G 6R8, Canada}

\author{B. Sullivan}
\affiliation{Iridian Spectral Technologies Ltd., Ottawa, ON K1G 6R8, Canada}

\author{K. Dolgaleva}
\affiliation{Department of Physics, University of Ottawa, Ottawa, ON K1N 6N5, Canada}
\affiliation{Max Planck Centre for Extreme and Quantum Photonics, Ottawa, ON K1N 6N5, Canada}
\affiliation{School of Electrical Engineering and Computer Science, University of Ottawa, Ottawa, ON K1N 6N5, Canada}

\author{R. W. Boyd}
\affiliation{Department of Physics, University of Ottawa, Ottawa, ON K1N 6N5, Canada}
\affiliation{Max Planck Centre for Extreme and Quantum Photonics, Ottawa, ON K1N 6N5, Canada}
\affiliation{School of Electrical Engineering and Computer Science, University of Ottawa, Ottawa, ON K1N 6N5, Canada}
\affiliation{University of Rochester, Rochester, NY 14627, USA}

\author{C. Genes}
\email{claudiu.genes@mpl.mpg.de}
\affiliation{Max Planck Centre for Extreme and Quantum Photonics, Ottawa, ON K1N 6N5, Canada}
\affiliation{Max Planck Institute for the Science of Light,
D-91058 Erlangen, Germany}
\affiliation{Department of Physics, Friedrich-Alexander-Universit\"{a}t Erlangen-N\"urnberg, D-91058 Erlangen, Germany}

\author{J.-M. M\'{e}nard}
\email{jmena22@uottawa.ca}
\affiliation{Department of Physics, University of Ottawa, Ottawa, ON K1N 6N5, Canada}
\affiliation{Max Planck Centre for Extreme and Quantum Photonics, Ottawa, ON K1N 6N5, Canada}
\date{\today}
\affiliation{School of Electrical Engineering and Computer Science, University of Ottawa, Ottawa, ON K1N 6N5, Canada}

\begin{abstract}
Physical and chemical properties of materials can be modified by a resonant optical mode. Such recent demonstrations have mostly relied on a planar cavity geometry, others have relied on a plasmonic resonator. However, the combination of these two device architectures have remained largely unexplored, especially in the context of maximizing light-matter interactions. Here, we investigate several schemes of electromagnetic field confinement aimed at facilitating the collective coupling of a localized photonic mode to molecular vibrations in the terahertz region. The key aspects are the use of metasurface plasmonic structures combined with standard Fabry-Perot configurations and the deposition of a thin layer of glucose, via a spray coating technique, within a tightly focused electromagnetic mode volume. More importantly, we demonstrate enhanced vacuum Rabi splittings reaching up to $200$\,GHz when combining plasmonic resonances, photonic cavity modes and low-energy molecular resonances. Furthermore, we demonstrate how a cavity mode can be utilized to enhance the zero-point electric field amplitude of a plasmonic resonator. Our study provides key insight into the design of polaritonic platforms with organic molecules to harvest the unique properties of hybrid light-matter states.
\end{abstract}


\maketitle

\section{Overview}

Strong light-matter interactions can modify fundamental properties of some physical systems leading to applications beyond fundamental science. In general, the light modes used in such experiments range from the visible into the region of mid-infrared. In the visible regime, where light couples to electronic transitions, experiments have e.g.~shown modifications of charge conductivity \cite{orgiu2015conductivity, hagenmuller2017charge},  photochemistry \cite{hutchinson2012modifying,munkhbat2018suppression} or single-molecule branching ratios \cite{cang2013giant, wang2019turning}. The condition for observing coherent exchanges between light and matter requires the coupling strength, also known as the vacuum Rabi splitting (VRS), to exceed all of the various loss rates in the system. To mitigate optical loss, high finesse optical resonators are utilized: standard designs in early pioneering efforts have made use of distributed Bragg reflector (DBR) microcavities with active media consisting of either an inorganic semiconductor quantum well design \cite{weisbuch1992observation} or organic semiconductors \cite{lidzey1998strong,lidzey1999room}.\\
%%%%%%%%%%%%%%%%%%%%%%%
\begin{figure}[!b]
\includegraphics[width=0.85\columnwidth]{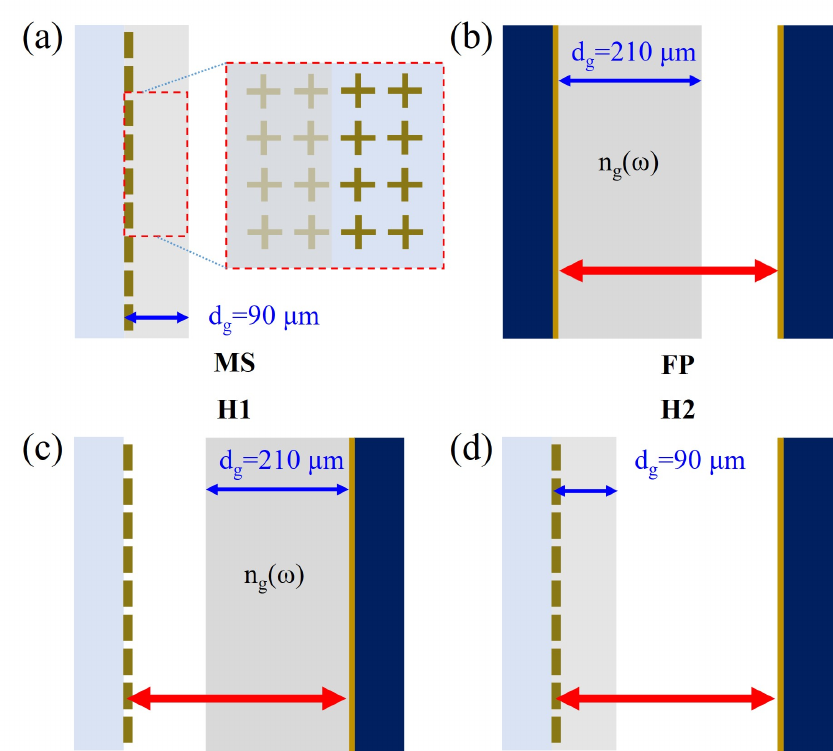}
\caption {\textbf{Hybrid architectures for light-matter coupling.} (a) Plasmonic metasurface (MS) as an array of cross-shaped metal elements coated with a $( 90\pm 10)$\,$\mu$m-thick glucose layer. (b) Fabry-Perot (FP) cavity where one mirror is coated with a $(210\pm 10)$\,$\mu$m-thick glucose layer. (c) and (d) Hybrid cavity designs in which one mirror of the FP cavity is replaced by a MS, and where the glucose layer either covers the mirror (H1) or the MS (H2). }
\label{fig1}
\end{figure}
%%%%%%%%%%%%%%%%%%%%%%%
\indent In the THz regime (wavelengths from $0.1$ to $1$\,mm), light can also directly couple to intra- and inter-molecular vibrational modes \cite{shalabney2015coherent,george2015liquid, jarc2022tunable}. This led to emerging applications in chemistry through the modification of ground state potential landscapes allowing for the manipulation of chemical reactions pathways and rates \cite{thomas2016ground,nagarajan2021chemistry, hirai2020modulation}. However, for the purpose of studying light-matter coupling involving far-infrared light and low-energy vibrational transitions of molecules, the DBR design becomes unpractical since most dielectrics are absorptive in this region and their required thickness exceeds the capacity of most nanofabrication equipment. In this context, other types of optical resonators have been demonstrated such as standard Fabry-Perot (FP) cavities with planar mirrors, where a VRS on the order of $68$\,GHz has been reached~\cite{damari2019strong}.\\
%%%%%%%%%%%%%%%%%%%%%%%

\indent We propose here alternative architectures where structural resonances in the THz regime are interfaced with vibrations in ensembles of organic molecular quantum emitters. On the electromagnetic side, this can refer to single plasmonic emitters/antennas~\cite{paulillo2016room,bayer2017terahertz}, metallic waveguides~\cite{TodorovY2009Strong,porer2012nonadiabatic, kaeek2021strong}, or metasurfaces (MSs)~\cite{dietze2011terahertz,scalari2012ultrastrong, benz2013strong}. On the matter side, we focus on glucose which
has a strong relevance to biological processes including
metabolism and photosynthesis. More specifically, we propose a set of four hybrid architectures illustrated in Fig.~\ref{fig1} to achieve a regime of strong light-matter interactions with a sharp vibrational resonance of glucose. In a first step, we demonstrate similar performance of both plasmonic MSs and standard FP cavities in achieving strong light-matter couplings. This is due to the stronger spatial confinement of a photonic mode allowed by the evanescent field of the MS which enables a reduction in the number of molecular emitters in comparison to the FP cavity. The combination between the two platforms can then lead to an improvement in the overall finesse of the system~\cite{cernotik2019cavity}.\\
\indent We focus on providing strong confinement of an electromagnetic mode around a vibrational transition of glucose at $1.43$\,THz which has a linewidth around $100$\,GHz, while ensuring a limited mode volume where the glucose layer is placed. In addition, full control of the resonance frequency and linewidth can be achieved by a tailored design of subwavelength arrays of metallic elements \cite{porterfield1994resonant,melo2008metal} with engineered periodicity and geometry of its constituent elements as a function of the surrounding media. This renders the MS as a frequency-selective mirror (analogous e.g., to subradiant optical mirrors realized with trapped atoms \cite{rui2020asubradiant}) which can be integrated in standard FP configurations to achieve better cavity finesse \cite{cernotik2019cavity}. A spray coating technique is used to deposit a layer of crystalline sugar directly on a MS, allowing for both the optimization of interaction with the plasmonic field as well as for the tuning of the plasmonic resonance \cite{gingras2020ultrafast}. The light-matter interaction strength, i.e., the vacuum Rabi splitting, is detected in the transmission of the hybrid system via a time-resolved THz spectroscopy technique which allows for the reconstruction of the full oscillating electric field. Rabi splittings in the range of $140$\,GHz proving strongly coupled light-matter dynamics are observed which are in good agreement with theoretical simulations based on the transfer matrix approach.\\

\section{Setup and approach}

MSs are constructed using a photolithography and metal evaporation process in which arrays of cross-shaped aluminium elements are fabricated onto a THz transparent 188\,$\mu$m-thick cyclo-olefin copolymer, commercially branded as Zeonor, with a measured index of refraction of $1.53$ (at $1$\,THz). The photolithography procedure utilizes a negative tone resist to etch the array design, followed by an aluminium deposition through direct current sputtering and a subsequent lift-off process. The dimensions of the metallic elements and the periodicity of the array, schematically shown in Fig.~\ref{fig1}(a), are optimized to achieve a sharp THz plasmonic resonance causing a distinctive dip in the spectral transmission. The four-fold symmetry of the cross-shaped elements is chosen to ensure a response insensitive to the polarization state of the THz radiation. A spray coating technique (see Appendix~\ref{Appendix C} for details) is used to deposit a glucose layer \cite{bargaoui2022comparative}.

In a first experiment, the layer is deposited directly on the MS as shown schematically in Fig.~\ref{fig1}(a).
%%%%%%%%%%%%%%%%%%%%%%%
%%%%%%%%%%%%%%%%%%%%%%%
\begin{figure}[t]
\includegraphics[width=0.9\columnwidth]{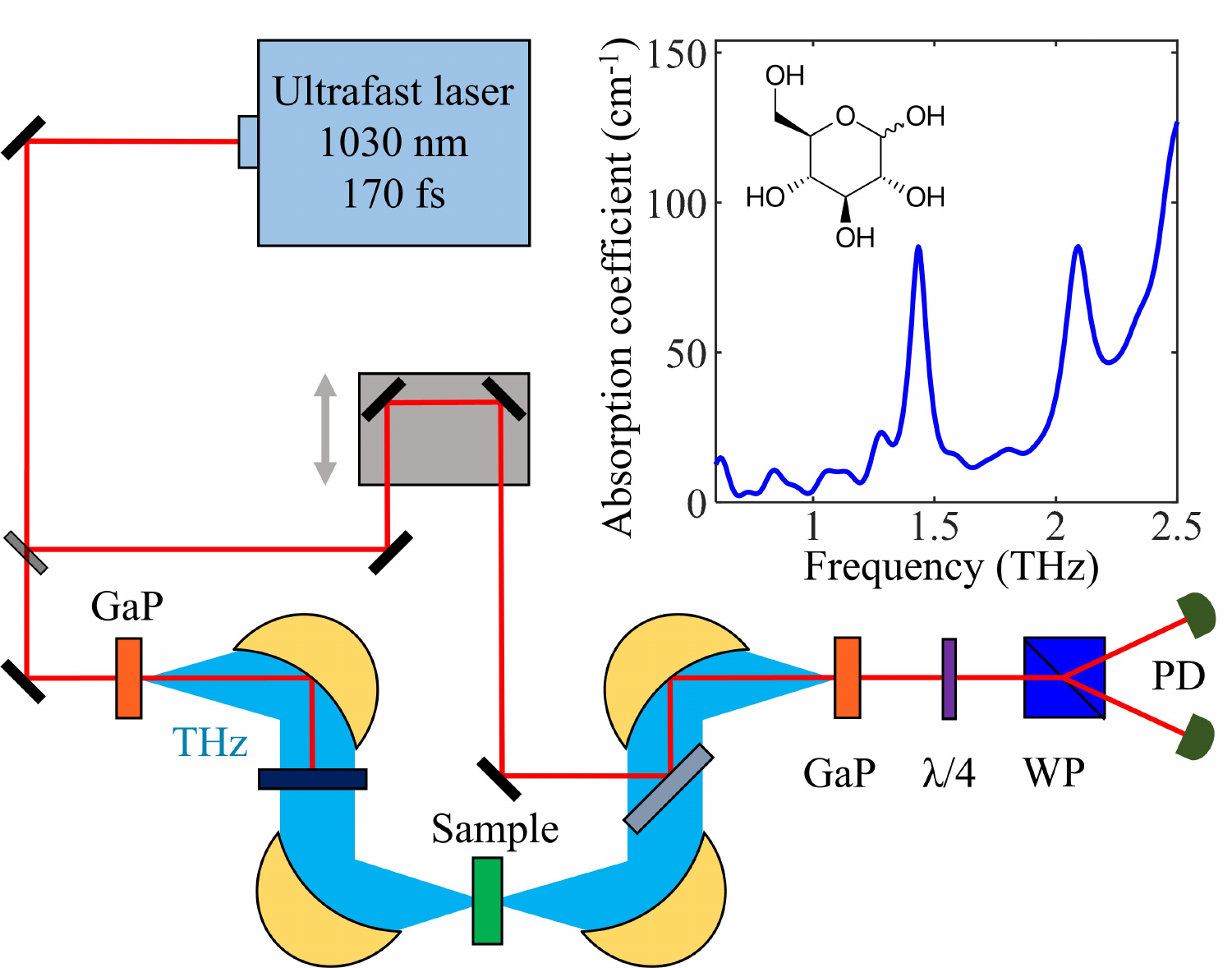}
\caption {\textbf{THz-TDS setup and the THz absorption coefficient of glucose.} Schematic of the THz time-resolved spectroscopy setup. An ultrafast laser source is used to generate THz through optical rectification in a GaP crystal. The detection process uses a partially reflected pulse from the same optical source to perform standard electro-optic sampling (EOS) inside another GaP crystal. In brief, the THz electric field is revealed by resolving THz-induced birefringence in the near-infrared gating pulse, which is monitored as a function of time delay with the THz pulse with a quarter-wave plate ($\lambda/4$), Wollaston prism (WP) and a pair of balanced photodiodes (PD). (inset) Absorption spectrum of a 300\,$\mu$m-thick glucose ($\mathrm{C}_6\mathrm{H}_{12}\mathrm{O}_6$) pellet measured with time-resolved THz spectroscopy and featuring a prominent vibrational resonance at $1.43$\,THz.}
\label{fig2}
\end{figure}
%%%%%%%%%%%%%%%%%%%%%%%
In a second experiment, two partially reflective mirrors ($R \approx 0.85$), fabricated by sputtering $9$\,nm of gold on a semiconductor substrate, form a FP cavity in the THz region (Fig.~\ref{fig1}(b)). The semiconductor substrate is $650\,\mu$m-thick undoped GaAs with a refractive index of $3.6$ (at $1$\,THz). In a third and fourth experiment, we explore hybrid architectures, schematically shown in Figs.~\ref{fig1}(c),(d), where one mirror of the FP cavity is replaced by a MS. We investigate the THz transmission response when a glucose layer is deposited on the planar mirror (H1) or on the MS (H2) with a thickness of $210~\mu$m and $90~\mu$m, respectively.

Characterization of the different architectures is performed with a THz time-domain spectroscopy (THz-TDS) technique, depicted in the setup shown in Fig.~\ref{fig2}. In brief, an ultrafast source operating at $1030$\;nm with a pulse duration of $170$\,fs and a repetition rate of $1.1$ MHz is split into a pump and gating beams. The pump generates a THz transient through optical rectification in a $2$\,mm-thick GaP crystal. A standard electro-optic sampling scheme resolves the THz pulse using another $2$\,mm-thick GaP crystal.
%The THz-induced birefringence, proportional to the local THz field, is resolved with polarization optics (a quarter wave place, a Wollaston prism and a pair of balanced photodiodes) as a translational stage (TS) changes the relative time delay between the two pulses. As a result, the full oscillating THz field can be resolved.

In order to observe a sizeable VRS, both the linewidths of the photonic device and that of the molecular ensemble have to be optimized. Saccharides, like glucose, are known to have a distinctive narrow radiative transition in the THz region originating from collective intermolecular vibrations due to the hydrogen bonding networks in the crystalline phase \cite{walther2003noncovalent}. The complex refractive index of glucose is measured with time-resolved THz spectroscopy. The inset of Fig.~\ref{fig2} shows the measured absorption spectrum with a prominent vibrational resonance at $1.43$\,THz and background absorption increasing towards higher frequencies. The real part of the refractive index is $n_g$  = 1.9 at $1.43$\ THz. We used these values to model the spectral transmission of the MS with the Lumerical FDTD solver \cite{lumericalref} and the response of the devices including a photonic cavity with a theoretical analysis based on the transfer matrix method (see Appendix~\ref{Appendix A} for details).

The description of the interaction between an electromagnetic mode (angular frequency $\omega_c$) and an ensemble of $N=\rho V$ (in a volume $V$ with number density $\rho$) molecular vibrational dipoles (angular vibrational frequency $\nu$) is given by the Tavis-Cummings Hamiltonian (setting $\hbar=1$)
\begin{equation}
\label{eq:taviscummings}
{\mathcal{H}}_\text{TC}=\omega_c \hat a^\dagger \hat a+\sum_j \nu \hat b_j^\dagger \hat b_j +\sum_j g_j(\hat a^\dagger \hat b_j+ \hat b_j^\dagger \hat a),
\end{equation}
where $\hat a$, $\hat b_j$ are the bosonic annihilation operators for the cavity mode and the $j$th molecular dipole, respectively. The position-dependent coupling $g_j=\mu \mathcal{E}f(\mathbf r_j)\boldsymbol\varepsilon_\mu^j\cdot \boldsymbol\varepsilon_c$ is given by the product of the dipole moment of the vibrational transition $\mu$, the zero point electric field amplitude $\mathcal{E}$ (inversely proportional to $1/\sqrt{V_\text{opt}}$ - where $V_\text{opt}$ is the cavity mode volume) and a spatial function $f(\mathbf r_j)$ evaluated at the position of the molecule $\mathbf r_j$. The unit vectors $\boldsymbol\varepsilon_\mu^j$ and $\boldsymbol\varepsilon_c$ account for the relative orientation between the molecular dipole and the cavity polarization, respectively. Assuming, for simplicity,  the case where all couplings are identical $g_j=g$, the Hamiltonian describes the coupling of a single collective bright mode $\hat B= \sum_j \hat b_j /\sqrt{N}$ to the cavity field with a collective coupling strength $g_N=g\sqrt{N}$. This is the so-called vacuum Rabi splitting (note that the splitting between the peaks is actually $2g_N$) and describes the strength of the light-matter coherent exchanges. In the limit where this rate dominates any loss processes, the system is said to be in the collective strong coupling regime. We remark that the Hamiltonian presented in Eq.~\eqref{eq:taviscummings} assumes the rotating wave approximation (RWA) which is valid as long as the collective coupling strength is much smaller than the transition frequency.

A very simple alternative to derive the Rabi splitting, even beyond the RWA, is to perform a linear response analysis via the transfer matrix approach. In this way, the splitting is obtained as the difference between the light-matter hybridized states (polaritons) in the transmission profile as a function of the incoming laser frequency (see Appendix~\ref{Appendix D}). \\
%%%%%%%%%%%%%%%%%%%%%%%
%%%%%%%%%%%%%%%%%%%%%%%
\begin{figure}[t]
\includegraphics[width=0.99\columnwidth]{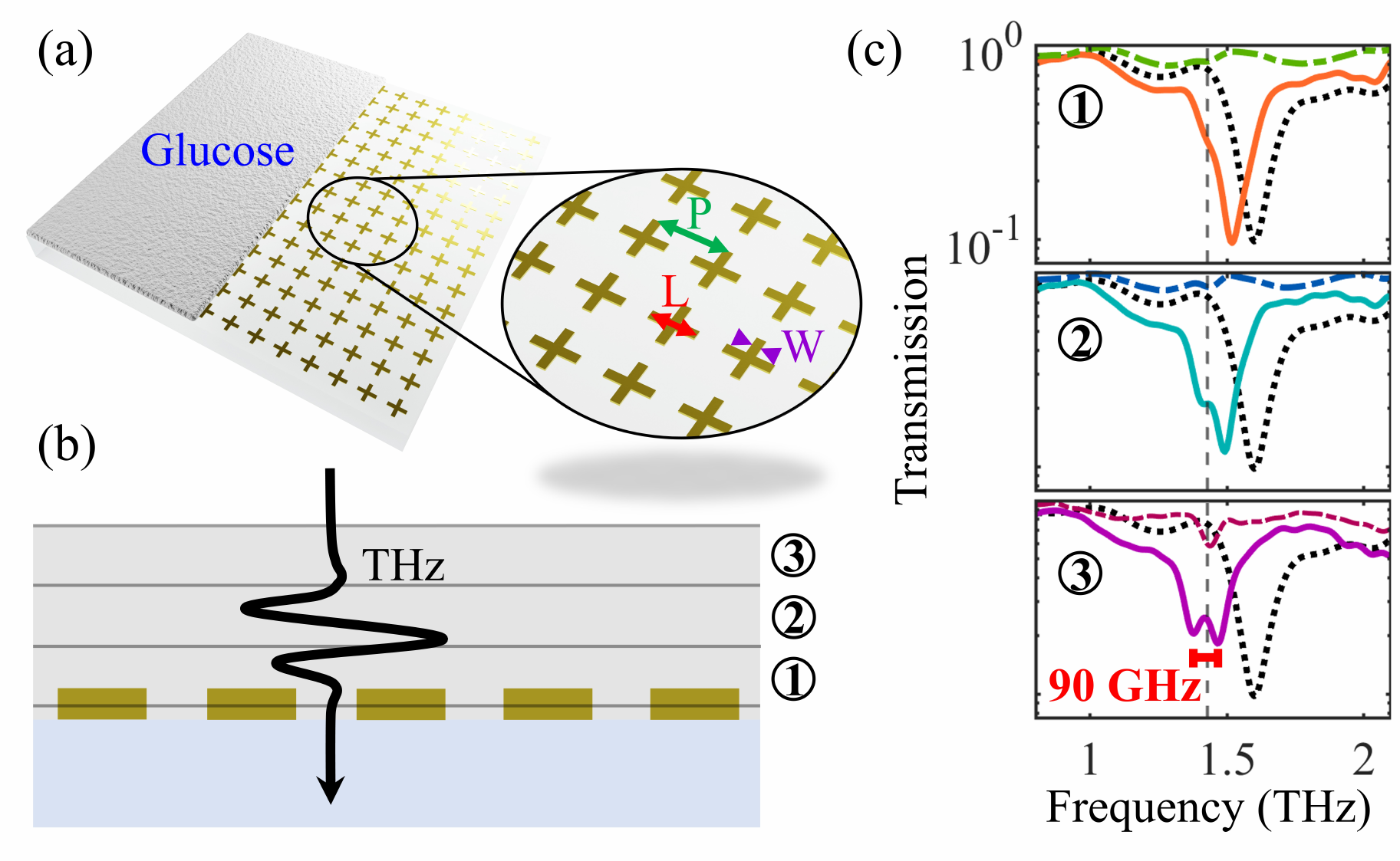}
\caption {\textbf{Strong light-matter coupling with a glucose-coated MS.} (a)  Schematic of a MS designed from an array of cross-shaped aluminium elements (shown in dark yellow color for clarity) with a glucose coating covering half the structure. The inset is a zoom-in defining the structural dimensions: the periodicity ($P$), cross arm length ($L$), and cross arm width ($W$), which are optimized to provide a narrow plasmonic resonance. (b) A cross-sectional schematic of the MS with three thicknesses of glucose layers:  (1) 30\,$\mu$m, (2) 60\,$\mu$m, and (3) 90\,$\mu$m, deposited with successive spray coating passes. (c) THz-TDS measurements taken of these three structures (lines) and the bare glucose layer on Zeonor (without the MS) (dashed lines). The transmission spectrum of the uncoated MS  is provided for comparison (black dots).}
\label{fig3}
\end{figure}
%%%%%%%%%%%%%%%%%%%%%%%
%%%%%%%%%%%%%%%%%%%%%%%%%%%%%%%%%%%%%%%%%%%%%%%%%%%%%%%%%%%%%%%%%%%%%%%%%%%%%%%%%%%%%%%%%%%%
%%%%%%%%%%%%%%%%%%%%%%%%%%%%%%%%%%%%%%%%%%%%%%%%%%%%%%%%%%%%%%%%%%%%%%%%%%%%%%%%%%%%%%%%%%%%

\section{Coupling with a metasurface resonator}

 The spectral transmission of a plasmonic MS does not only depend on the geometry and periodicity of its metallic elements, but also on the background media surrounding its interface. The design shown in Fig.~\ref{fig1}(a) presents a Zeonor substrate underneath the MS and glucose or air on the top surface. The structure is optimized to allow strong coupling between the molecular resonance at $1.43$\,THz and a resonant plasmonic mode. A rendered depiction of the sample is shown in Fig.~\ref{fig3}(a) with an inset zoom-in that defines the MS geometry in terms of periodicity ($P$), cross arm length ($L$) and cross arm width ($W$). Since the coupling strength scales with $\sqrt{N/V_\text{opt}}=\sqrt{\rho V/V_\text{opt}}$, the glucose layer must ideally fill up the plasmonic mode volume to ensure that the largest number of emitters are coupled with the electromagnetic mode. Further addition of glucose will not increase the coupling, which is obviously limited by the density $\rho$, but instead only brings detrimental effects due to the increased absorption. We therefore iteratively deposit thin layers of glucose (height $d_g$, cross section $A$) while monitoring the transmission properties. A cross sectional diagram of the coated MS is illustrated in Fig.~\ref{fig3}(b) where three coating thicknesses are picked to showcase the improvement of the light-matter interaction.

The complex frequency-dependent transmission coefficient for the hybridized MS-glucose system can be obtained from transfer matrix theory as (neglecting effects of the substrate)
\begin{align}
t_\mathrm{MS}(\omega)=\frac{e^{i\omega d_g n_g(\omega)/c}}{1-i\zeta_\mathrm{hyb}(\omega)},
\end{align}
with $c$ the speed of light and $\zeta_\mathrm{hyb}$ describes the so-called polarizability of the hybridized MS. The expression for $\zeta_\mathrm{hyb}$ derived in Appendix~\ref{Appendix E} shows that the Rabi splitting increases with the thickness of the glucose layer and eventually saturates once the thickness exceeds the plasmonic mode volume as characterized by the penetration depth $z_0$. The collective coupling can then be expressed as $g_N= g_0\sqrt{\rho A z_0(1-e^{-2d_g/z_0})/2}$ where $g_0$ is the average coupling strength of a single emitter directly at the metasurface. 

\indent The measured intensity transmission of the coated MS is plotted in Fig.~\ref{fig3}(c). The bare MS (black dashed curve) initially has a plasmonic resonance at $1.6$\;THz corresponding to the narrow spectral dip in transmission. As we gradually deposit layers of glucose on the surface, we observe a red-shift of the resonance to (1) $1.52$\;THz,  (2) $1.49$\,THz, and (3) $1.43$\,THz (colored solid lines) with contributions of the Rabi splitting, leading to a maximum peak separation of $90$\,GHz at close to zero detuning. The response of an equal thickness coating of glucose on the substrate in an area with no MS elements is also plotted (colored dashed lines). The shift of the resonance can be understood as a gradual increase in the effective index of the medium on the top of the MS as we increase the thickness of the glucose layer. Consistent to the theoretical prediction, we also have observed that further increasing the volume of glucose to exceed the plasmonic mode volume will no longer yield a redshift or an improvement in the Rabi splitting, but instead only increased absorption by glucose. The vibrational resonance frequency of the targeted glucose mode is plotted for clarity (vertical dashed line). \\
%%%%%%%%%%%%%%%%%%%%%%%
%%%%%%%%%%%%%%%%%%%%%%%
\begin{figure*}[t]
\includegraphics[width=0.96\textwidth]{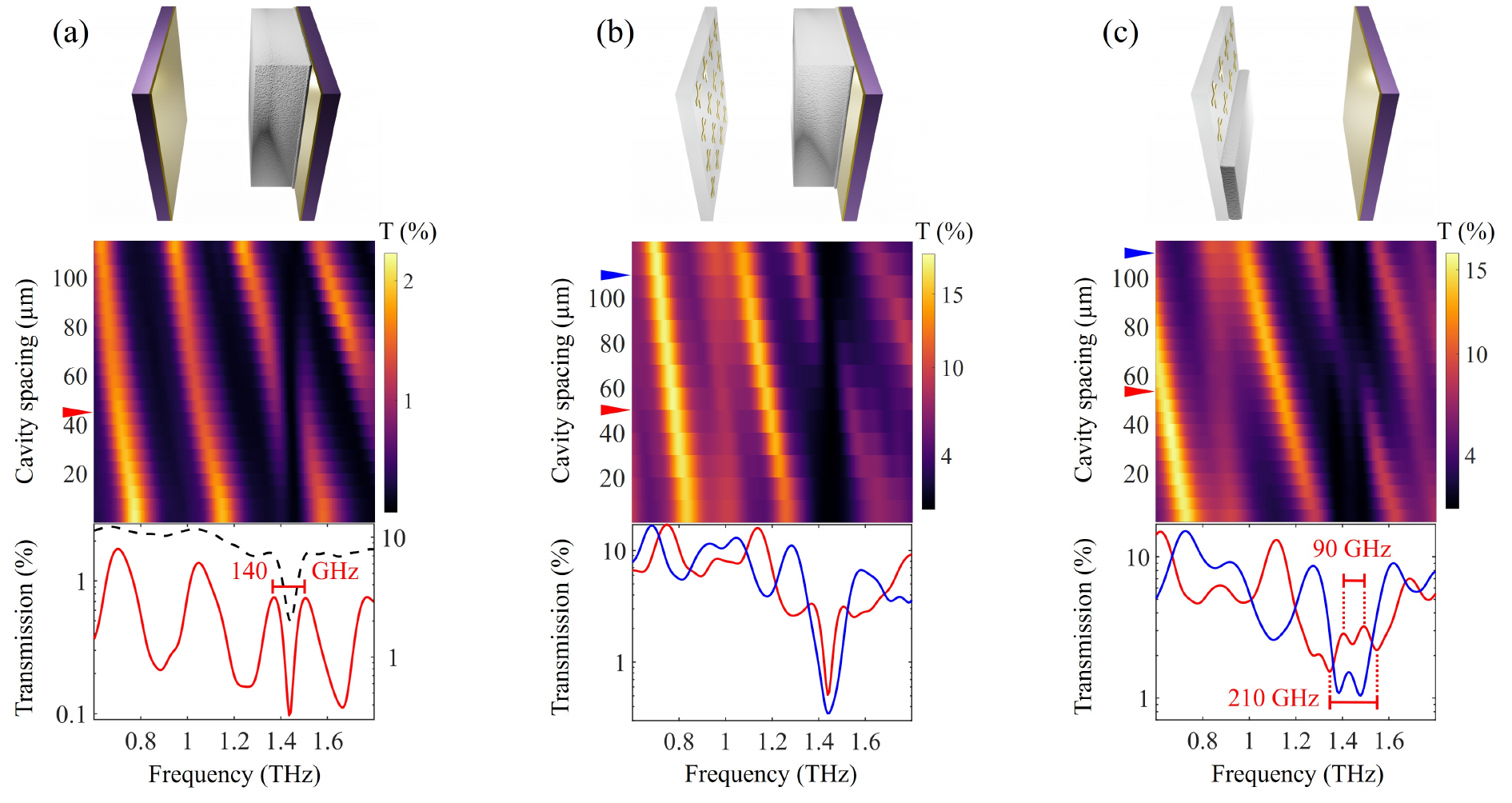}
\caption {\textbf{Strong light-matter coupling with FP and hybrid cavity architectures.} Transmission spectra for (a) standard FP cavity, (b) H1 and (c) H2. Top to bottom row shows the schematics of the cavity architectures, the 2D transmission as a function of relative cavity spacing and frequency as well as cross sections through the transmission profile. In the plots at the bottom, the dashed black curve in (a) shows the transmission of a glucose-coated gold mirror while the red and blue curves show the transmission for different cavity lengths (on-resonance/off-resonance) as indicated by the arrows in the density plot above. The H2 architecture leads to an enhanced polaritonic response, showing a substantially broader Rabi splitting when a cavity mode is overlapping with the polaritons of the coupled MS. 
}
\label{fig4}
\end{figure*}
%%%%%%%%%%%%%%%%%%%%%%%

\section{Coupling with a standard Fabry-Perot cavity}

 A rendered depiction of a FP cavity filled with glucose is shown in the top row of Fig.~\ref{fig4}(a). The flat mirror cavity is created from partial mirrors formed by sputtered gold on GaAs. One of the mirrors is coated with glucose using the aforementioned spray coating technique. THz-TDS measurements are taken at $5\,\mu$m increments of spacing between the cavity mirrors. This effective spacing is given by $d_\text{eff}(\omega) = d_\text{air} + d_g n_g(\omega)$, where $d_\text{air}$ is the air space between the mirrors and $n_g(\omega)$ is the refractive index of glucose.
Analytically, one can deduce the expression for the transmission
\begin{align}
t_\mathrm{FP}(\omega)=\frac{ e^{-i\omega d_\mathrm{eff}(\omega)/c}}{\zeta^2+(1-i\zeta)^2 e^{-2i\omega d_\mathrm{eff}(\omega)/c}},
\end{align}
where $\zeta$ describes the polarizability of the gold mirrors which we assume to be frequency-independent. This expression is then used to fit the experimental results (see Appendix~\ref{Appendix F}).

Experimental results are presented in Fig.~\ref{fig4}(a). The middle row shows a scan of the transmission as a function of frequency for varying cavity spacings. Around the $1.43$\,THz resonance of glucose, one of the cavity modes displays an anti-crossing behavior that is characteristic of strong light-matter coupling. The corresponding Rabi splitting at around $140$\,GHz is observed in the bottom row of Fig.~\ref{fig4}(a). \\

\section{Coupling with hybrid cavity architectures}

 While strong coupling can be achieved in both setups described above, integrating MSs with flat mirrors might bring a few advantages. One stems from the convolution of the MS's narrow frequency response with the cavity transmission window to design sharper resonances~\cite{cernotik2019cavity} and thus sharper polaritonic peaks. Secondly, the coupling between the three resonances involving glucose, MS and cavity, leads to a richer polaritonic physics beyond the standard upper and lower polaritons typical of strong coupling experiments.
We therefore test two scenarios, depicted in Figs.~\ref{fig4}(b) and (c).

%%%%%%%%%%%%%%%%%%%%%%%
\begin{figure*}[t]
\includegraphics[width=0.75\textwidth]{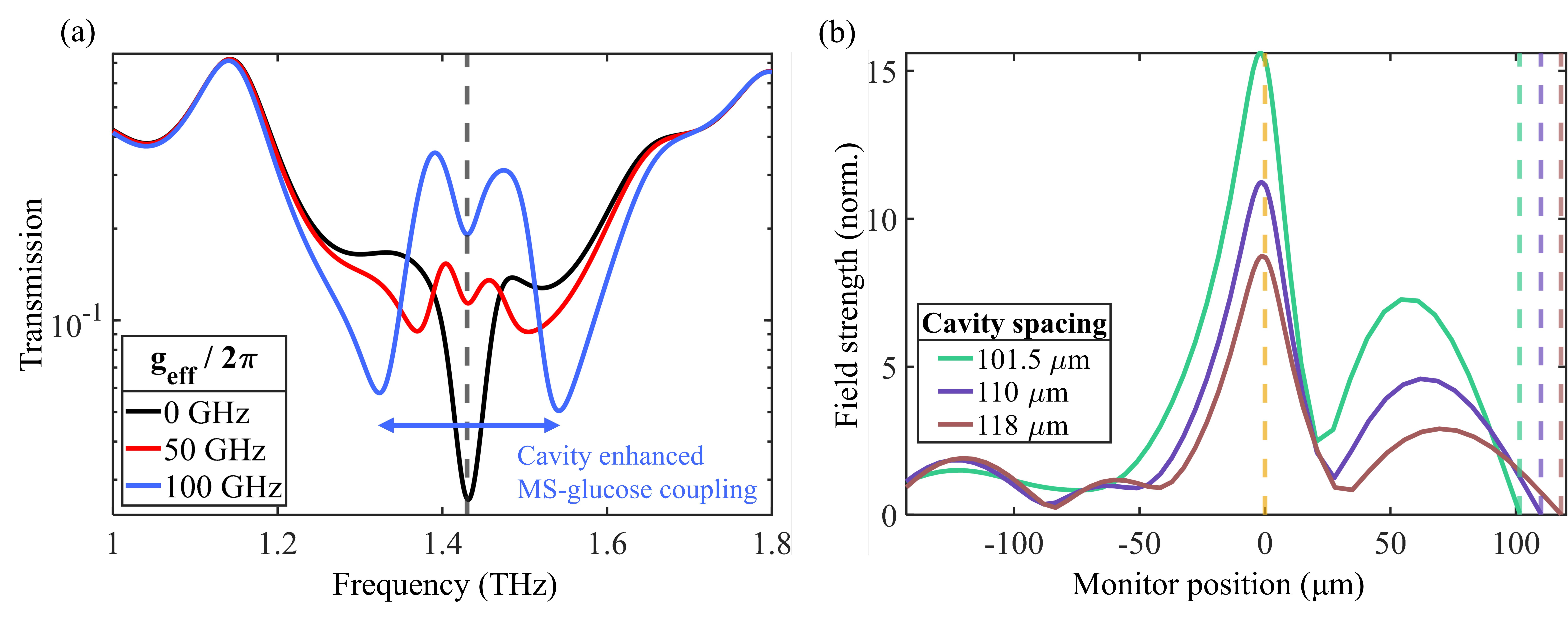}
\caption {\textbf{Transfer matrix and FDTD simulations of the hybrid cavity.} (a) Transfer matrix simulation of cavity transmission (on-resonance) for different glucose-MS coupling strengths $g_\mathrm{eff}$ and $d_g = 100\,\mu$m. The arrows indicate the enhanced Rabi splitting of the MS-glucose interaction due to the cavity configuration. The vertical dashed line shows the location of the glucose resonance. When $g_\mathrm{eff} > 0$, the glucose-MS polaritons are dominant and the transmission result resembles experimental observations of the H2 configuration. Otherwise, when $g_\mathrm{eff}  = 0$, only the cavity-glucose interaction remains, and the transmission result resembles the H1 configuration.  (b) A FDTD investigation of an empty hybrid cavity. The position of the mirror, relative to the array interface ($0\, \mu$m, orange dashed line), determines the cavity spacing and thus the cavity resonance. Reducing the cavity spacing (from brown to purple to green) allows one to bring the cavity mode in resonance to the MS mode. The electric field magnitude within ($> 0\, \mu$m) and outside ($< 0\, \mu$m) of the cavity is monitored in one dimension and normalized to the incident field. The green curve shows the field profile obtained when the plasmonic and cavity modes are resonant to each other. The corresponding vertical dashed lines show the positions of the mirror. The enhancement of the field amplitude is consistent with the modified Rabi splitting showcased in (a).}
\label{fig5}
\end{figure*}
%%%%%%%%%%%%%%%%%%%%%%%
The H1 architecture involves an uncoated MS with a resonance frequency of $1.45$\,THz merged with a glucose coated flat mirror with a coating thickness of 210\,$\mu$m. Results are depicted in the middle row density plot of Fig.~\ref{fig4}(b). In the density plot, one can see an anti-crossing region forming around the $1.43$\,THz vibrational resonance of glucose. Additionally, as a cavity mode shifts towards the MS/vibrational resonance, the linewidth of the cavity mode can be shown to narrow slightly. The MS is functioning as a strongly frequency-dependent mirror with an approximately Lorentzian response. The bottom row of Fig.~\ref{fig4}(b) compares the transmission characteristics of the H1 architecture when a cavity mode is overlapped with the MS/vibrational resonance (on-resonance) versus when a cavity mode is spectrally far from the MS/vibrational resonance (off-resonance). We observe a narrowing of the resonance as well as the formation of polariton peaks when the cavity mode is resonant with the MS/vibrational resonance. The splitting between the peaks in the on-resonance transmission is measured to be $135$\,THz, comparable to the splitting observed in the case of the flat mirror cavity with the same glucose coating thickness.\\\\

The architecture of H2, depicted in the top row of Fig.~\ref{fig4}(c), is particularly suited for the study of complex polaritonic systems as it couples the glucose resonance to both the cavity-delocalized mode and to the MS resonance. This is observed in the middle plot as an interesting anti-crossing behavior. Distinct dark anti-crossing regions can be seen when the cavity mode is spectrally far from the polaritonic modes of the MS/glucose. This off-resonant case corresponds to the transmission response of just the coated MS. In contrast, when the mirror spacing is set so that the cavity mode begins to interact with the polaritons, the anti-crossing region morphs into three dark regions. This interaction can be observed more clearly by looking at slices of the transmission when the cavity mode is on-resonance (overlapping with the polaritons) versus off-resonance (spectrally far from the polaritons). Transmission spectra for the on- and off-resonant cases are plotted in the bottom row of Fig.~\ref{fig4}(c). We can observe how in the on-resonant case, the response of the H2 architecture shows an enhancement of the MS-glucose interaction which leads to an effectively larger Rabi splitting of $200$\,GHz for the MS-glucose polaritons. 
\\

\section{The nature of the hybrid cavity enhacement}

To better understand the competition between the cavity-glucose and MS-glucose coupling for the H2 cavity architecture, we show in Fig.~\ref{fig5}(a) a plot of the transfer matrix simulation results with increasing MS-glucose coupling strength $g_\mathrm{eff}$, while keeping all other parameters fixed. When geff is small, the cavity-glucose interaction can be seen from the formation of two weak polariton peaks symmetrically shifted from the dominant glucose resonance. In fact, when $g_\mathrm{eff} = 0$, we retrieve the transmission response observed for the H1 architecture (black curve). When increasing the MS-glucose coupling, which is the case for H2 architecture, two dips appear, corresponding to the MS-glucose polaritons. Most importantly, the exhibited splitting is now roughly a factor of two larger than the case of coupling only to the MS. This enhancement can be traced back to the fact that the hybrid architecture leads to an increase in the zero-point electric field amplitude of the intra-cavity mode around the position where the glucose is added. To prove this point, we incorporate FDTD simulations, shown in Fig.~\ref{fig5}(b), of an empty hybrid cavity. The electric field strength is monitored in one-dimension along the MS-cavity axis and normalized to the incident field. The design consists of an infinitely periodic planar array of metallic crosses with a perfectly reflecting mirror plane at a distance (cavity spacing) above. In the plot, the orange dashed line indicates the position of the MS and the green, purple, and brown dashed lines indicate the various positions of the mirror planes. Three different cavity spacings are compared to showcase when a cavity mode is completely overlapped (green curve), partially overlapped (purple curve), and not overlapped (brown curve) with a cavity mode. The electric field is monitored at the resonance frequency of the plasmonic array. Positive monitor positions are within the cavity and negative monitor positions are outside of the cavity. The resultant field profiles obtained from these simulations clearly show that the field strength at the MS interface of a hybrid cavity can be significantly enhanced (for example a factor around $1.8$ close to the interface and even larger further away from it) when a cavity mode is resonant to a plasmonic mode. As aforementioned, the coupling strength, $g$, is linearly proportional to the local electric field given by the product of the zero-point electric field amplitude and the spatial function of the mode supported by the photonic resonator.

\section{Conclusion}

In summary, we have shown strong coupling between a vibrational resonance of glucose and a plasmonic MS mode in the THz regime where powerful spectroscopic methods based on the THz-TDS technique can be exploited for system characterization. Furthermore, we suggest that this work can open avenues in designing light-matter interfaces based on hybrid architectures. We have shown experimentally, and with support from analytics tested against numerical simulations, the emergence of strong light-matter coupling in a variety of geometries, ranging from the standard FP cavity to more complex hybrid configurations interfacing a MS with a planar mirror. We have demonstrated the enhancement of light-matter coupling strength brought on by the hybrid cavity design and directly connected it to the increase in the zero-point electric field amplitude stemming from the interference between the MS evanescent field and the standing wave field of the cavity at the location of the glucose layer. Further investigations in this direction hold the promise of identifying mechanisms for the design of higher-finesse cavities as well as providing platforms for the exploration of the richer physics of multi-polariton systems.\\

\noindent \textbf{Data availability.} The experimental data that is plotted in this study is available upon reasonable request from the corresponding authors.\\

\noindent \textbf{Code availability.} The code used for experimental analysis and the transfer matrix calculations is available upon reasonable request from the corresponding authors.\\

\noindent \textbf{Acknowledgments.} We  acknowledge funding from the Natural Sciences and Engineering Research Council of Canada (NSERC) Strategic Partnership Program (STPGP/ 521619-2018), NSERC Discovery funding program (RGPIN-2016-04797) and Canada Foundation for Innovation (CFI) (Project Number 35269). C.\,G. and M.\,R.~acknowledge financial support from the Max Planck Society and the Deutsche Forschungsgemeinschaft (DFG, German Research Foundation) -- Project-ID 429529648 -- TRR 306 QuCoLiMa
(``Quantum Cooperativity of Light and Matter''). \\

\bibliography{RefsHybrid}

\onecolumngrid

\setcounter{figure}{0}
\setcounter{table}{0}
\makeatletter
\renewcommand{\thefigure}{A\arabic{figure}}
\renewcommand{\bibnumfmt}[1]{[A#1]}

\newpage
\appendix

\section{Response of plasmonic MS}
\label{Appendix A}

We provide here a quick review of the reflection and transmission properties of two-dimensional MSs as widely covered in the literature, largely following Ref.~\cite{garcia2007colloquium}. We consider a square array of (ideal, pointlike) dipoles with resonance frequency $\omega_0=ck_0=2\pi c/\lambda_0$ (wavenumber $k_0$,  resonance wavelength $\lambda_0$), located in the $xy$ plane at positions $\mathbf{r}_j$ with lattice constant $a$. Each scatterer responds with its induced dipole moment $\mathbf{p}_j=\boldsymbol\alpha_\mathrm{p} \mathbf{E}(\mathbf{r}_j)$, determined by its polarizability tensor $\boldsymbol\alpha_\mathrm{p}$ and the electric field acting on it and produces a field of $\mathbf E_\text{dip}^j(\mathbf{R})=\mathbf G (\mathbf{R}-\mathbf{r}_j)\mathbf{p}_j$ at position  $\mathbf{R}$ governed by the free space Green's tensor (evaluated at the dipole resonance) \cite{novotny2006principles, garcia2007colloquium}
\begin{align}
\label{dipdip}
\mathbf G (\mathbf{R})= (k_0^2+\nabla\otimes\nabla)\frac{e^{ik_0|\mathbf{R}|}}{|\mathbf{R}|}.
\end{align}
The self-consistent dipole moment of the $j$th element is given by
\begin{align}
\mathbf p_j =\boldsymbol\alpha_\mathrm{p}\left[\mathbf E_\mathrm{in}(\mathbf r_j)+\sum_{j'\neq j}\mathbf G (\mathbf r_j-\mathbf r_{j'})\mathbf p_{j'}\right],
\end{align}
which has contributions from the incident field as well as the re-radiated fields from all surrounding dipoles.
Let us consider a quasi-infinite array under normal, plane wave illumination $\mathbf E_\mathrm{in} (\mathbf R)=\mathbf E_\mathrm{in} e^{ik_\ell z}$ with wavenumber $k_\ell$. In this case, all dipoles will respond identically, i.e., $\mathbf p_j =\mathbf p$, yielding a straightforward solution to the above equation as
\begin{align}
\mathbf p = \frac{\mathbf E_\mathrm{in}}{\boldsymbol\alpha_\mathrm{p}^{-1}-\widetilde{\mathbf G}(0)},
\end{align}
where $\widetilde{\mathbf G}(0)=\sum_{j\neq 0} \mathbf G (\mathbf r_j-\mathbf r_0)$ corresponds to the sum of the Green's tensor over all lattice vectors (w.r.t.~some arbitrary lattice vector $\mathbf r_0$). More generally, for non-normal incidence of the incoming light, this extends to the lattice Fourier transform of the Green's tensor evaluated at the in-plane quasimomentum $\mathbf q$: $\widetilde{\mathbf G} (\mathbf q)=\sum_{j\neq 0} \mathbf G(\mathbf r_j) e^{-i\mathbf q\cdot\mathbf r_j}$. The total electric field is obtained as
\begin{align}
\mathbf E (\mathbf R)=\mathbf E_\mathrm{in}(\mathbf R)+\mathbf p \sum_j \mathbf G (\mathbf R-\mathbf r_j),
\end{align}
requiring the estimate of the sum of all dipole-radiated fields. This sum over the real-space lattice can be turned into a sum over the reciprocal lattice via Poisson's summation formula. For subwavelength arrays $a<\lambda_0$, this expression greatly simplifies as in the far-field only a single term survives while all others are evanescent. In addition only considering a single polarization component (e.g., all dipoles pointing along $x$), the far field along the $z$ direction is obtained as (assuming $k_0\approx k_\ell$)
\begin{align}
E^{\mathrm{far}}(z)=E_\mathrm{in} \left[e^{ik_\ell z}+ r_m e^{ik_\ell |z|}\right],
\end{align}
with the complex reflection coefficient of the MS
\begin{align}
r_m=\frac{ 2\pi i k_0/a^2}{1/\alpha_\mathrm{p}^{xx}-\widetilde G_{xx} (0)},
\end{align}
while the complex transmission is obtained as $t_m=1+r_m$. The frequency-dependent reflectivity can be rewritten in a more intuitive Lorentzian form  as
\begin{align}
\label{eq:reflectivity}
r_m(\omega)=-i\frac{\gamma+\widetilde\gamma (0)}{(\omega-\omega_0-\widetilde\Omega(0))+i(\gamma+\widetilde\gamma (0))},
\end{align}
where we defined the MS-induced frequency shift $\widetilde\Omega (0)$ and decay rate modification $\widetilde\gamma (0)$ as
\begin{subequations}
\begin{align}
\label{eq:decayrate}
\widetilde\gamma (0) &=\frac{3\gamma}{2k_0^3}  \mathrm{Im}\left[\widetilde G_{xx}(0)\right]  =\gamma\frac{3}{4\pi}\left(\frac{\lambda_0}{a}\right)^2-\gamma,\\
\widetilde\Omega (0) &= -\frac{3\gamma}{2k_0^3}  \mathrm{Re}\left[\widetilde G_{xx}(0)\right].
\end{align}
\end{subequations}
The last step in Eq.~\eqref{eq:decayrate} holds for subwavelength arrays $a<\lambda_0$. Here, we have assumed non-absorbing dipoles with a polarizability of (in cgs units)
\begin{align}
\alpha_\mathrm{p}^{xx} (\omega)=-\frac{3}{16\pi^3}\lambda_0^3\frac{\gamma}{(\omega-\omega_0)+i\gamma},
\end{align}
where $\gamma$ is the radiative linewidth of a single dipole. In the following, we denote the MS linewidth as $\gamma_m=\gamma+\widetilde\gamma (0)$ and the MS resonance frequency as $\omega_m=\omega_0+\widetilde\Omega (0)$. Note that the above treatment is in principle only valid for particle sizes much smaller than the wavelength of the incident light. For larger particle sizes, electrodynamic corrections have to be taken into account (for details see e.g.,~Refs.~\cite{jensen1999electrodynamics, binalam2021ultra}).

\section{THz time-domain spectroscopy}
\label{Appendix B}

All experimental measurements taken for this work in the THz region are accomplished with a time-domain spectroscopy technique (THz-TDS) using the setup depicted in Fig.~2 of the main text. Briefly, an ultrafast source operating at $1030$\;nm with a pulse duration of $170$\,fs and a repetition rate of $1.1$ MHz is split into a pump and gating beams. The pump generates a THz transient through optical rectification in a $2$\,mm-thick GaP crystal. A standard electro-optic sampling scheme resolves the THz pulse using another $2$\,mm-thick GaP crystal. The THz-induced birefringence, proportional to the local THz field, is resolved with polarization optics (a quarter wave place, a Wollaston prism and a pair of balanced photodiodes) as a translational stage (TS) changes the relative time delay between the two pulses. As a result, the full oscillating THz field can be resolved.

\section{Sample fabrication}
\label{Appendix C}

\subsection{MS fabrication}

MS designs utilized for this work were fabricated with a photolithography process using a photomask. The $150\,$nm thick aluminium cross array was deposited atop a 188\,$\mu$m-thick cyclo-olefin copolymer, commercially branded as Zeonor, with a measured index of refraction of $1.53$ (at $1$\,THz). The Zeonor substrate was selected due to its dispersionless quality and THz transparency. The photolithography procedure utilizes a negative tone resist to etch the array design, followed by an aluminium deposition through direct current sputtering and a subsequent lift-off process. Two MS designs are shown in the results of this work and defined in terms of the cross periodicity ($P$), arm length ($L$), and arm width ($W$). The glucose coated MS, with a coating thickness of $(90\pm 10)$\,$\mu$m, utilized for the results of Fig.~3 and H2 of Fig.~4 in the main text is defined by the geometry $P/L/W = 110.25/63/9.54\,\mu$m and has a resonance frequency of 1.6 THz. For the results of H1 of Fig.~4 in the main text, a MS with geometry $P/L/W = 124.25/71/10.76\,\mu$m, with a resonance frequency at 1.45 THz, is fabricated.

\subsection{Cavity mirror fabrication}

Planar mirror samples for the cavity architectures described in this work were fabricated by sputtering $9\,$nm of gold on a $650\,\mu$m-thick undoped GaAs substrate with a refractive index of $3.6$ (at $1$\,THz). One mirror sample was then coated with a $(210\pm 10)$\,$\mu$m-thick glucose layer and utilized for the FP and H1 cavity architectures shown in Fig.~4 of the main text. Another mirror sample which was not coated is utilized to produce the FP and H2 cavity architectures shown in Fig.~4 of the main text.

\subsection{Glucose spray coating technique}

\begin{figure}[h]
\includegraphics[width=0.7\columnwidth]{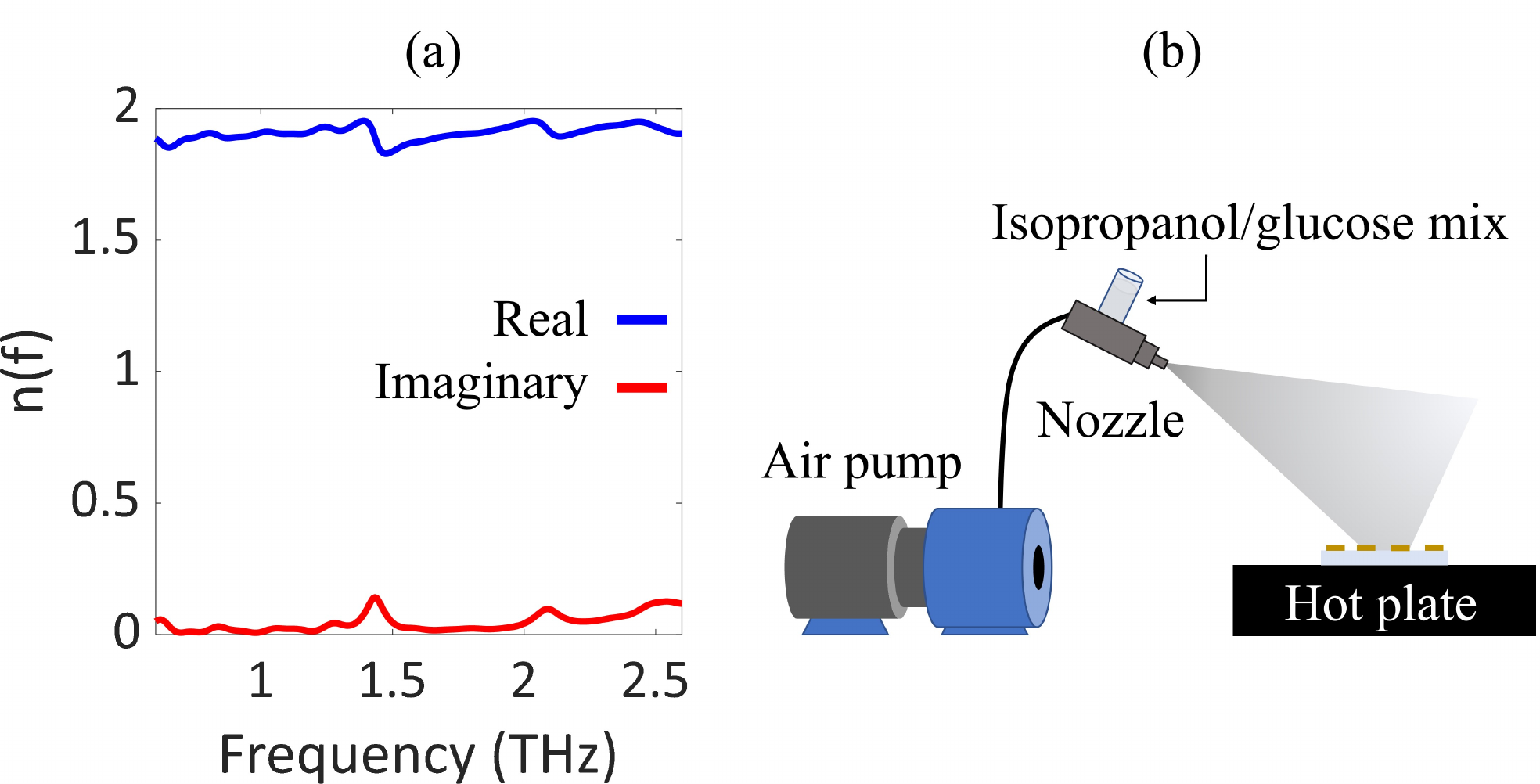}
\caption {(a) Extracted refractive index measurement of a glucose pellet with $300$\,$\mu$m thickness created using a hydraulic press. Glucose has an index of 1.9 (real part) at the targeted 1.43 THz vibrational mode. (b) Cartoon schematic of the spray coating technique. A suspension of sonicated glucose powder in isopropanol is air sprayed onto a sample which is placed onto a hotplate. The temperature of the hotplate is set to allow isopropanol to rapidly evaporate, leaving a layer of solid-state glucose on the sample.}
\label{figS1}
\end{figure}

$\alpha$-D-Glucose was acquired (MilliporeSigma) in a solid-state powder form. The THz refractive index of glucose was measured from a pellet sample created utilizing a hydraulic press to achieve a pellet diameter of 500\,mm with a $300$\,$\mu$m thickness. The real and imaginary parts of the extracted index are shown in Fig.~\ref{figS1}(a). While pellets are commonly used in standard FP cavity experiments, we found that they lacked robustness for experiments involving MSs due to pellet fragility, thickness/diameter scaling, and the inability to ensure a strong concentration of glucose between the plasmonic elements of our arrays.

\indent We instead use a spray coating technique to place a large concentration of glucose molecules at the interface of the plasmonic array, depicted in the cartoon schematic of Fig.~\ref{figS1}(b). The vibrational resonances of glucose observed at 1.43 and 2.09\,THz are prominent when glucose is in solid-state form. All glucose-coated MSs and planar mirrors produced for this work were coated using this spray coating technique.

\noindent Placing many solid-state glucose molecules near the plasmonic interface of the MS requires the crystalline particle size of glucose to be reduced; we achieve this condition utilizing an ultrasonic bath. First, we prepare a mixture of glucose powder in isopropanol, which is non-polar and has a low evaporation temperature. The mixture is then sonicated to produce a stable suspension of glucose in isopropanol with crystalline sizes of $\sim 10$\,$\mu$m. The extracted suspension of fine glucose particulates are loaded into an air spray gun. The sample to be coated is placed onto a hot plate with a temperature set above the boiling point of isopropanol (82.5\,\degree C) to facilitate the rapid evaporation of the solvent. A few sprays of the mixture onto the substrate and subsequent drying results in a relatively uniform layer of solid-state glucose on the substrate. To increase the layer thickness, the cycle of spray coating and drying is repeated. The thickness of spray coated glucose onto the samples is probed using optical microscopy.

\section{FDTD simulation of the MS field distribution}
\label{Appendix D}

\begin{figure}[h]
\includegraphics[width=0.55\columnwidth]{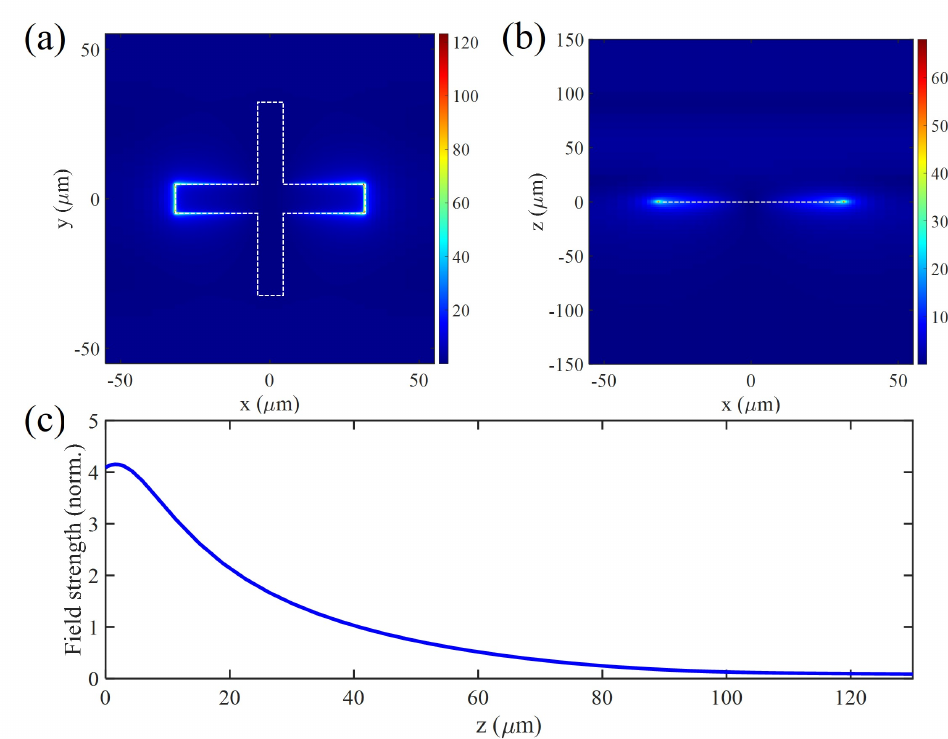}
\caption {FDTD simulations of field distribution at the air/substrate/MS interface  (shown is a single plasmonic element) (a)  in the $xy$ plane ($z=0$\,$\mu$m) and (b) in the $xz$ plane ($y=0$\,$\mu$m) spanning 150\,$\mu$m above and below the interface. (c) Evanescent field strength
monitored along z at a position ($x = 41.5$\,$\mu$m, $y = 0$\,$\mu$m), which is chosen because it shows a distribution representative of the
overall field contained within the $xy$ plane above the metasurface. }
\label{figS2}
\end{figure}

To investigate the evanescent nature  of the plasmonic mode volume of the MS, we utilize a FDTD simulation to monitor the electric field distribution. Using the geometry of the cross shown in Fig.~3 of the manuscript with periodic boundary conditions, we can simulate the response of the MS. Figs.~\ref{figS2}(a) and (b) show the spatial field distribution of a cross element at the plasmonic resonance frequency in the $xy$ and $xz$ planes, respectively. The incident pulse injected into the simulated structure is linearly polarized along the $x$ axis. We then investigate the evanescent field
strength in the glucose along the $z$-axis. Fig.~\ref{figS2}(c) shows the absolute field calculated at the position ($x = 41.5$\,$\mu$m, $y = 0$\,$\mu$m), which is representative of the overall values obtained numerically at different xy positions above the
metasurface.

\section{Transfer matrix theory}
\label{Appendix E}

Transfer matrix theory is a simple yet powerful method to obtain the transmission and reflection coefficients and thereby the spectral response of layered systems \cite{born1999principles, mackay2020transfermatrix}. The method connects the field amplitudes of left ($-$) and right ($+$) traveling waves on the left of the structure $(E_L^-, E_L^+)^\top$  to the left- and right traveling waves on the right $(E_R^-, E_R^+)^\top$ via the transfer matrix $\mathbf T$ such that
\begin{align}
\begin{pmatrix}E_L^-\\E_L^+\end{pmatrix}=\mathbf T \begin{pmatrix}E_R^-\\E_R^+\end{pmatrix}.
\end{align}
Assuming a unit input field entering the system from the left $E_L^+=1$,  and no input field from the right, $E_R^-=0$, the transmission coefficient  is simply obtained as the inverse of the last entry of the transfer matrix $t=1/T_{22}$. Furthermore, the complex transmission and reflection amplitudes can always be connected as $t=1+r$. The free propagation through a dispersive medium with index of refraction $n(\omega)$ and length $d$ is described by the matrix
\begin{align}
\mathbf T_f=
\begin{pmatrix} e^{i\omega n(\omega) d/c}& 0\\0 & e^{-i\omega n(\omega) d/c}\end{pmatrix},
\end{align}
while the action of a thin mirror or MS is described by
\begin{align}
\mathbf T_m=
\begin{pmatrix} 1+i\zeta_m (\omega)& i\zeta_m (\omega)\\-i\zeta_m (\omega) & 1-i\zeta_m (\omega)\end{pmatrix}.
\end{align}
Here, the quantity $\zeta_m (\omega)=-ir_m (\omega)/t_m (\omega)$ is called the polarizability of the mirror which describes the ratio between the complex reflection and transmission coefficients. This is not to be confused with the polarizability from section \ref{Appendix A}. While this is a strongly frequency-dependent quantity for the MS, we assume the polarizability of the gold mirrors to be constant within the considered frequency range. Assuming a Lorentzian reflectivity for the MS (compare with Eq.~\eqref{eq:reflectivity}), given by $r_m(\omega)=-i\gamma_m/[(\omega-\omega_m)+i(\gamma_m+\gamma_l)]$ and including an additional nonradiative loss channel $\gamma_l$, the polarizability of the MS simply expresses as
\begin{align}
\zeta_m(\omega) = \frac{\gamma_m}{(\omega_m-\omega)-i\gamma_l}.
\end{align}
 In addition, the Fresnel losses at the interface between two layers $j$ and $j+1$ can be taken into account by the transfer matrix
\begin{align}
\mathbf T_{j,j+1}=
\frac{1}{t_{j,j+1}}\begin{pmatrix} 1& r_{j,j+1}\\r_{j,j+1} & 1\end{pmatrix},
\end{align}
with the standard Fresnel reflection and transmission coefficients
\begin{align}
r_{j,j+1}=\frac{n_j-n_{j+1}}{n_j+n_{j+1}},\qquad t_{j,j+1}=\frac{2n_j}{n_j+n_{j+1}}.
\end{align}
For the purpose of modeling spray-coated glucose, we take a Lorentz-Drude model for the refractive index around the resonance at $\nu=2\pi \cdot 1.43$\,GHz (note that $\omega$ and $\nu$ are angular frequencies)
\begin{align}
n_g(\omega)^2=\varepsilon_g (\omega)=\varepsilon_{g,\mathrm{back}}+\frac{\alpha_g\nu^2}{\nu^2-\omega^2-i\gamma_g \omega},
\end{align}
with background permittivity $\varepsilon_{g,\mathrm{back}}=3.61$, oscillator strength $\alpha_g=0.028$, and linewidth $\gamma_g=2\pi\cdot 85$\,GHz.  

\begin{figure}[h]
\includegraphics[width=0.55\columnwidth]{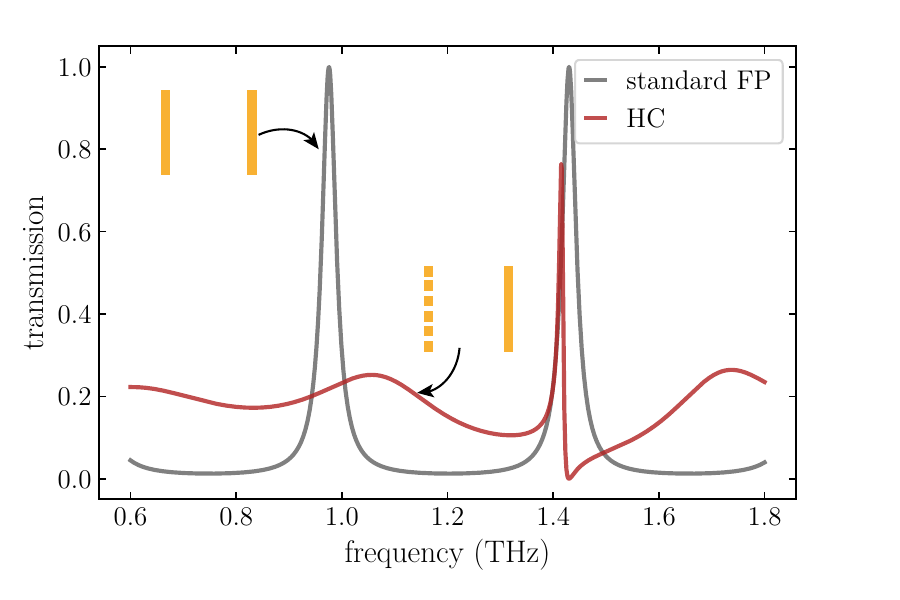}
\caption {Comparison of cavity transmission functions for (a) standard FP cavity with $\zeta=2$ and (b) hybrid cavity design where one of the mirrors is replaced by a MS with linewidth $\gamma_m=2\pi\cdot 50$\,GHz and resonance frequency $\omega_m=2\pi\cdot 1.43$\,THz. The cavity length for both architectures is $348$\,$\mu$m.}
\label{figS3}
\end{figure}

The total transmission matrix is then simply obtained by multiplying the matrices of the individual elements. In Fig.~\ref{figS3} we illustrate the potential advantages of hybrid cavity designs by comparing the transmission of a standard FP cavity to that of a hybrid architecture where the left mirror is replaced by an ideal, lossless MS. The hybrid architecture leads to a filtering out of all other resonances while simultaneously showing a reduction in linewidth around the resonance of interest (with a Fano-like profile and a zero of transmission located directly at the MS resonance).

\section{Coupled-dipoles approach for glucose-MS coupling}
\label{Appendix F}

\begin{figure}[t]
\includegraphics[width=0.55\columnwidth]{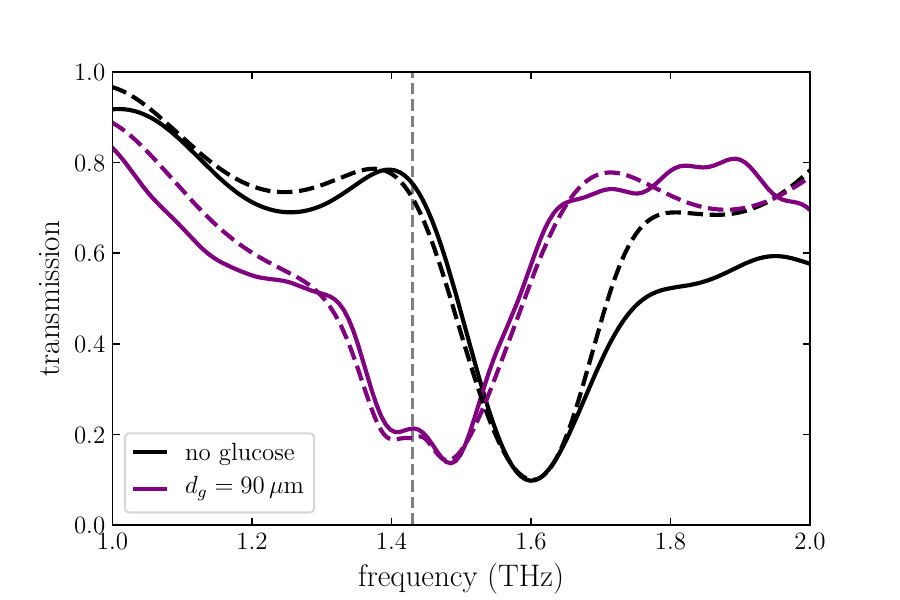}
\caption {Comparison of transmission between uncoated (black, $d_g=0\,\mu$m) and glucose-coated (purple, $d_g=90\,\mu$m) MS. The solid curves show the experimental results while the dashed curves show the transfer matrix fits which also take into account the effects of the Zeonor substrate. The MS is fitted with $\gamma_m=2\pi\cdot 100$\,GHz, $\gamma_l=2\pi\cdot 52$\,GHz, and we assume a coupling strength of $g_\mathrm{eff}=2\pi\cdot 45$\,GHz for $d_g=90\,\mu$m. The vertical dashed line shows the location of the glucose resonance.}
\label{FigS4}
\end{figure}

While the transfer matrix method describes the forward and backward propagation of plane waves through the structure, it does not per se include the near-field coupling of the glucose layer to the evanescent field of the MS. This coupling can be included by a coupled-dipoles model. To this end, one assumes that each vibrational dipole couples to the MS with a distance dependent rate $g_j\equiv g(\mathbf r_j)$. From the Tavis-Cummings Hamiltonian (Eq.~(1) of the manuscript), supplemented with the proper decay terms, one can then derive equations of motion for the amplitudes of the electric field of the MS and the $j$th molecular dipole in the glucose layer
\begin{subequations}
\begin{align}
\dot\beta_m&=-(\gamma_m+\gamma_l+i\omega_m)\beta_m -i\sum_j g_j\beta_g^{(j)}+\eta (t),\\
\dot\beta_g^{(j)}&=-(\gamma_g+i\nu)\beta_g^{(j)} -i g_j\beta_m,
\end{align}
\end{subequations}
where $\eta (t)$ describes some arbitrary input field probing the response of the MS and $\gamma_g$ is the linewidth of the glucose resonance. Solving this set of equations in Fourier domain, one can see that the coupling to the glucose gives rise to a modified MS response
\begin{align}
\beta_m(\omega)=\frac{\eta (\omega)}{i(\omega_m-\omega)+\gamma_m+\gamma_l+\frac{\sum_j g_j^2}{\gamma_g+i(\nu-\omega)}},
\end{align}
and therefore also a renormalized polarizability of the MS
\begin{align}
\label{eq:renormalized}
\zeta_\mathrm{hyb}(\omega)=\frac{\gamma_m}{(\omega_m-\omega)-i(\gamma_m+\gamma_l)-i\frac{\sum_j g_j^2}{\gamma_g+i(\nu-\omega)}}.
\end{align}
Assuming the field to fall off exponentially from the MS, we can write $g_j=g(z_j)=g_0 e^{-z_j/z_0}$  where $z_0$ is the penetration depth of the field into the glucose layer ($\approx 15\,\mu$m) and $g_0$ is the coupling strength directly at the MS (considering that we already averaged over the field distribution in the $xy$ plane). Considering a total number of $N$ molecular emitters coupling to the MS ($N=A d_g \rho$, with $A$ the cross section of the glucose layer, $d_g$ the thickness of the glucose layer and $\rho$ the density of molecules), the sum over the coupling strength can be turned into an integral
\begin{align}
\sum_j g_j^2=A\rho g_0^2\int_0^{d_g} dz \,e^{-2z/z_0}=g_0^2\frac{N_{z_0}}{2}\left(1-e^{-2d_g/z_0}\right),
\end{align}
where $N_{z_0}$ describes the number of emitters within the mode volume $N_{z_0}=A z_0\rho$. From this, we can define the effective coupling strength (Rabi frequency) as $g_\mathrm{eff}(d_g)=\sqrt{\sum_j g_j^2}$. In Fig.~\ref{FigS4} we make use of Eq.~\eqref{eq:renormalized} to fit the transfer matrix result to the experimentally measured transmission (parameters in caption).

\section{Experiment - Transfer matrix theory comparison}
\label{Appendix G}

\begin{figure}[h]
\includegraphics[width=0.78\columnwidth]{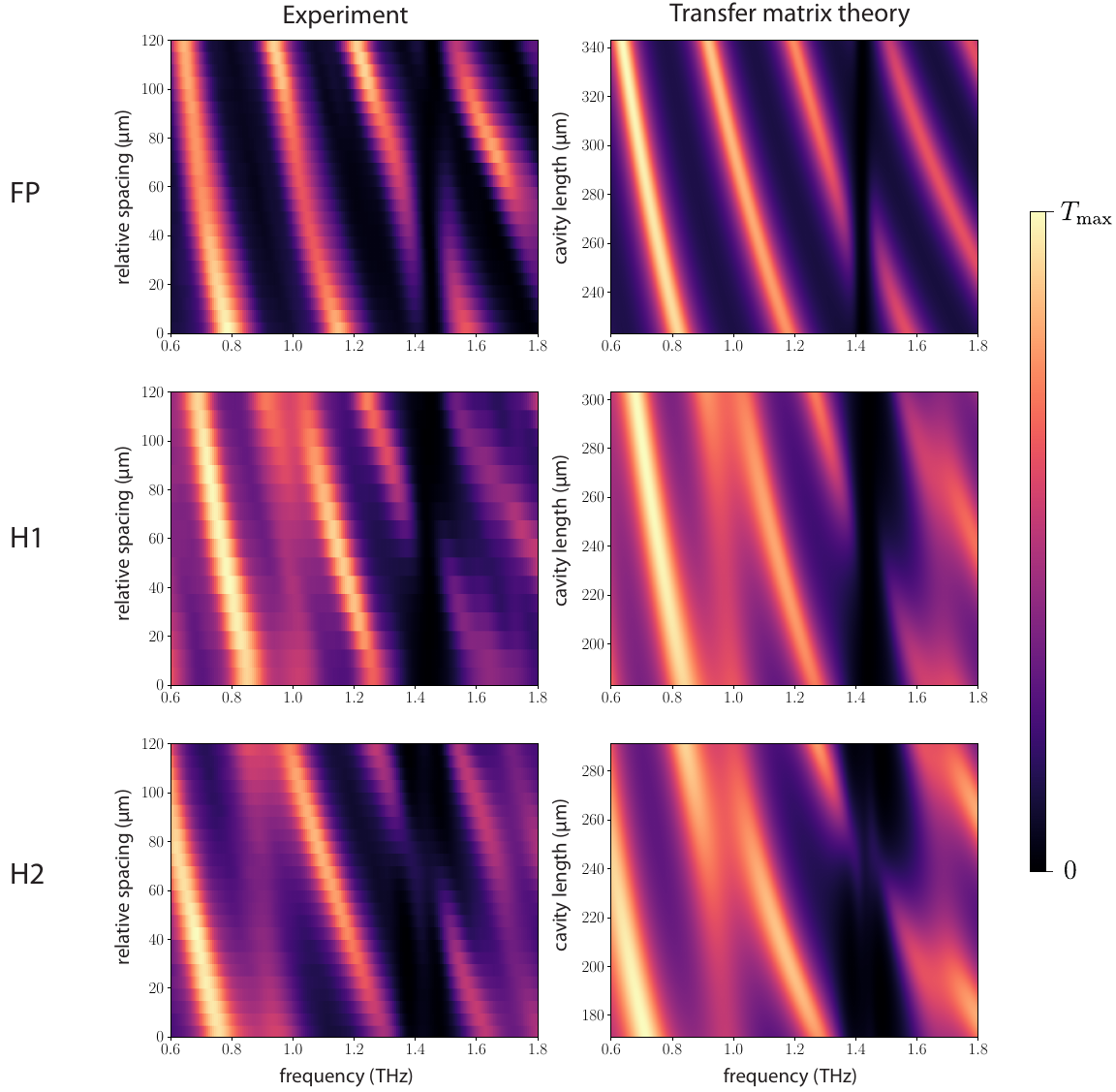}
\caption {Comparison between measured transmission spectra (left column) and results obtained from transfer matrix theory (right column)  for the different cavity architectures.  Each density plot is normalized to the maximum transmission intensity $T_\mathrm{max}$. The MS is fitted with $\gamma_m=2\pi\cdot 100$\,GHz, $\gamma_l=2\pi\cdot 52$\,GHz, $\omega_m=2\pi\cdot 1.43$\,THz while for the gold mirrors we assume $\zeta=0.9$. For H2, we assumed an effective MS-glucose coupling strength of $g_\mathrm{eff}=2\pi\cdot45$\,GHz. The horizontal resonances at approximately $1\,\mathrm{THz}$ for the H1 and H2 architectures stem from back reflection in the Zeonor substrate. We neglected back reflection from the GaAs substrate as this is not captured by the THz-TDS measurement.}
\label{figS5}
\end{figure}

In Fig.~\ref{figS5} we show a comparison between the experimental results and transfer matrix simulations. For the hybrid architectures, we used the MS fit described in Section~\ref{Appendix E}. More details can be found in the figure caption.

\clearpage

\end{document}